\documentclass[%
 pra, twocolumn, 10pt,
superscriptaddress,
 amsmath,amssymb,
 aps
]{revtex4-1}

\usepackage{graphicx}
\usepackage{dcolumn}
\usepackage{bm}
\usepackage{colortbl}
\usepackage{upgreek}

\begin{document}

\preprint{APS/123-QED}

\title{The resonant behavior of a single plasmonic helix}

\author{K. H\"{o}flich}
\email[Corresponding author: ]{katja.hoeflich@helmholtz-berlin.de}
\affiliation{%
Helmholtz Zentrum f\"{u}r Materialien und Energie Berlin,\\ Hahn-Meitner-Platz 1, D -- 14109 Berlin, Germany
}%
 \affiliation{Max Planck Institute for the Science of Light}
\author{Th. Feichtner}%
 \affiliation{%
 Nano-Optics \& Biophotonics Group, Department of Experimental Physics 5, R\"{o}ntgen Research Center for Complex Material Research (RCCM), Physics Institute, University of W\"{u}rzburg, Am Hubland, D-97074 W\"{u}rzburg, Germany
}%
\author{E. Hansj\"{u}rgen}%
\affiliation{%
 Institute of Physics and Center of Interface Science, Carl von Ossietzky Universit\"{a}t Oldenburg, D-26129 Oldenburg, Germany
}%
\author{C. Haverkamp}
\affiliation{%
Helmholtz Zentrum f\"{u}r Materialien und Energie Berlin,\\ Hahn-Meitner-Platz 1, D -- 14109 Berlin, Germany
}%
 \affiliation{Max Planck Institute for the Science of Light}

\author{H. Kollmann}
\author{C. Lienau}
\author{M. Silies}
\affiliation{%
 Institute of Physics and Center of Interface Science, Carl von Ossietzky Universit\"{a}t Oldenburg, D-26129 Oldenburg, Germany
}%

\date{\today}

\begin{abstract}
Chiral plasmonic nanostructures will be of increasing importance for future applications in the field of nano optics and metamaterials. Their sensitivity to incident circularly polarized light in combination with the ability of extreme electromagnetic field localization renders them ideal candidates for chiral sensing and for all-optical information processing. Here, the resonant modes of single plasmonic helices are investigated. We find that a single plasmonic helix can be efficiently excited with circularly polarized light of both equal and opposite handedness relative to that of the helix. An analytic model provides resonance conditions matching the results of full-field modeling. The underlying geometric considerations explain the mechanism of excitation and deliver quantitative design rules for plasmonic helices being resonant in a desired wavelength range. Based on the developed analytical design tool, single silver helices were fabricated and optically characterized. They show the expected strong chiroptical response to both handednesses in the targeted visible range. With a value of 0.45 the experimentally realized dissymmetry factor is the largest obtained for single plasmonic helices in the visible range up to now. 
\end{abstract}

\maketitle


\section{Introduction}

Chirality or handedness is a geometric property of all objects that cannot be superimposed onto their mirror image. The paradigm of a chiral object is represented by the helix. In nature, helices find their manifestations e.g.~in the geometry of amino acids, DNA or climbing plants. In optics, the electric field vector of left (right) circularly polarized light follows a left-handed (right-handed) helix upon propagation for a fixed point in time~\footnote{According to Hecht~\cite{Hecht2005} RCP light is defined by an electric field vector rotating clockwise at a fixed point in space when looking towards the light source. This is equivalent to a right-handed helix formed by the electric field vector at a fixed time. }. Chiral light can be used as a probe for chiral matter~\cite{Barron:2009:MLS:1816638, Tang2010,Schaferling2014}. However, the interaction of circularly polarized light with naturally occurring chiral objects is inherently weak. To significantly enhance light matter interaction in the visible range, metal nanostructures can be employed~\cite{Kreibig,Maier2007,Collins2017}. These can exhibit collective oscillations of their free electron gas under visible light incidence, so-called plasmons-polaritons, which in turn lead to extreme electromagnetic field concentrations. Accordingly, plasmonic helices constitute the nearly perfect choice in terms of maximized chiroptical interaction~\cite{Fernandez-Corbaton2016}.

Apart from their strong chiroptical response, helical plasmonic antennas show a complex resonance behavior which was not fully understood until now. For the radio-frequency regime an empirical description of the reception and emission behavior of helical antennas is known~\cite{balanis1982}\footnote{The plasmonic helix works similar to a helical antenna in axial or endfire mode.}. Towards the visible range the mismatch between free space and plasmon wavelength on a metallic wire leads to an additional degree of freedom. Furthermore, losses start to play a crucial role. Earlier studies on plasmonic helices attributed observed spectral features to a superposition of the respective split-ring eigenmodes of a 1-turn helix with Bragg resonances when moving to more than one turn~\cite{Gansel2009}. This explains the large asymmetry in the transmission of gold helical arrays, which strongly scatter and absorb incident light matching the handedness of the helix geometry and are transparent to light of opposite handedness. Numerical investigations on single silver helices proposed an effective dipole length onto the wire to describe the resonant modes~\cite{Zhang2011}.  Recently, a single loop gold helix was investigated in detail concerning the excitation of the fundamental resonance~\cite{Wozniak2018}. A multipole analysis revealed that the one-loop helix can be described as a point-like chiral molecule with nearly parallel electric and magnetic dipole moments on resonance~\cite{Barron:2009:MLS:1816638}. All these studies concentrate on the opto-chiral response for equal handedness of helix and incident field~\footnote{Zhang et al.~employ an inverted definition of LCP and RCP light.}. 

Another important observation is the possible zero crossing of the dissymmetry factor. The dissymmetry factor g normalizes the difference in extinction between left-circularly and right-circularly polarized (LCP and RCP) light to the total extinction (cf. Supplementary Information SI.). For the case of molecules it is known, that light of opposite handedness compared to the molecule handedness may also lead to efficient excitation~\cite{Barron:2009:MLS:1816638}. Indeed, such a spectral response was experimentally observed before for the case of chiral 'plasmonic molecules' of various geometries~\cite{Menzel2010,Yin2013,Cui2014,Mark2013,Frank2013,Kang2015,Qu2017,Lee2018}. This zero crossing can be qualitatively described in terms of a coupled oscillating dipoles within the Born-Kuhn model~\cite{Kuhn1930,Yin2013}. Similarly, ensembles of plasmonic helices from different metals showed a distinct asymmetry in transmission ($T_\text{RCP}- T_\text{LCP}$) with zero crossing~\cite{Gibbs2013,Esposito2014, Kosters2017}. However, the Born-Kuhn model cannot be transferred to the case of plasmonic helices for which  multiple reflections of plasmon polariton modes and retardation effects cause a wealth of higher order resonances and a rich spectral behavior. Hence, the full description of all types of resonant helix modes remains an open issue.

\section{Analytical model}
\begin{figure}[t]
\centering
\includegraphics[width=0.7\linewidth]{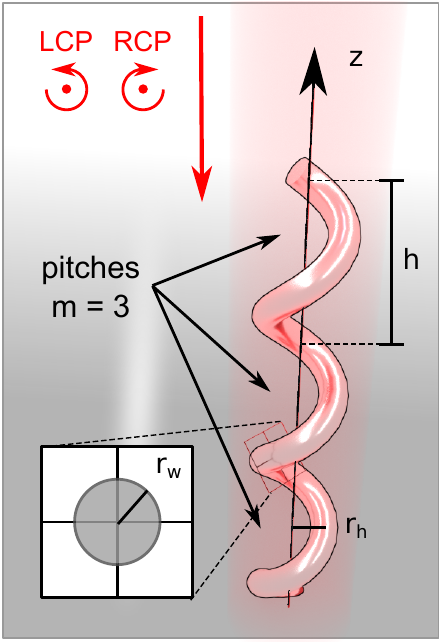}
\caption{\textbf{Geometry of a helical plasmonic antenna.} A single plasmonic helix of m turns is illuminated by circularly polarized light propagating along the helix axis.
}
\label{fig:geom}
\end{figure}
\begin{figure*}[t]
\centering
\includegraphics[width=0.75\linewidth]{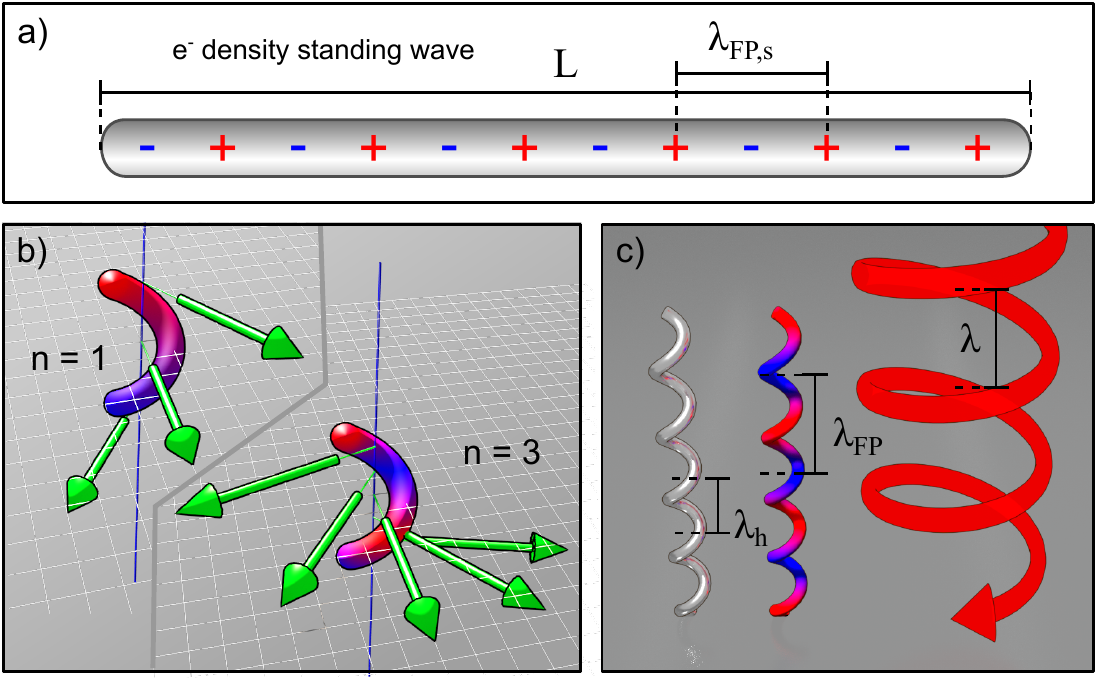}
\caption{\textbf{Resonant behavior of a helical plasmonic antenna.} a) For negligible near-field coupling between neighbouring turns, the eigenmodes of the antenna coincide with those of a straight plasmonic wire of length L being the wire length of the helix. The charge displacement associated with the corresponding plasmonic Fabry-Perot resonances are indicated by the + and - signs. b) The overlap of the incident light vectors with the mode patterns on the helical geometry determines the excitation efficiency of the respective eigenmode. c) Finally, the ratios between three different scales determined the actual excitation efficiency.
}
\label{fig:model}
\end{figure*}

Figure~\ref{fig:geom} depicts the geometrical setting of this work. Circularly polarized light is incident on a plasmonic helix. The plane wave propagates along the helix axis. To understand the chiroptical response of such a helical nanostructure let us start with a straight metallic wire. Upon excitation by an external electromagnetic field of a certain frequency an excitation of a standing wave may occur. Optically, the associated charge density oscillation can be understood as plasmonic Fabry-Perot mode~\cite{Vogelgesang} (cf. Figure~\ref{fig:model}a)). Standing wave patterns evolve when the wavelength meets the following condition:
\begin{equation}
\lambda_\text{eff}=\frac{2}{n}L = \frac{2m}{n} \sqrt{4\pi^2r_{\text{h}}^2 + h^2}, \hspace{2mm} n= 1,2, ...\hspace{2mm}.
\label{eq:FB}
\end{equation}
The wire length $L$ is given by the arc length of the helix with radius $r_{\text{h}}$ and $m$ turns of height $h$. The number of electric field antinodes is counted by $n$, denoting the mode order. For straight wires oblique incidence breaks the symmetry such that even and odd modes can be excited. The same argument applies for the helical geometry.

The effective wavelength $\lambda_\text{eff}$  of the plasmonic mode is reduced in comparison to the exciting free space wavelength $\lambda$ and follows a nearly linear scaling behavior~\cite{Novotny2007}:
\begin{equation}
\lambda_\text{eff} (\lambda) = 2\pi/\gamma(\lambda) - 4r_\text{wire}/n.
\label{eq:lambdaeff}
\end{equation} 
In the visible range, the propagation constant $\gamma(\lambda)$  is inversely proportional to the incident wavelength $\lambda$. The correction term involves the wire radius $r_\text{wire}$ and the order of the mode $n$.

Together, equations \eqref{eq:FB} and \eqref{eq:lambdaeff} allow us to calculate the excitation wavelengths for Fabry-Perot resonances of different orders $n$ on linear plasmonic antennas. These modes can be regarded as discrete set of eigenmodes for the helical geometry as long as the interaction between the helix turns is negligible (see Supplementary Information SIII.). 

Now let us tackle the geometry dependent efficiency of excitation of the eigenmodes. A mode is efficiently excited  if the spatial distribution of the incident electric field vectors resembles that of the local mode pattern on the helix surface~\cite{Hecht2005,Feichtner2017}. For the case of a single-turn helix this is true when the handedness of the light matches the handedness of the structure. As shown in Figure ~\ref{fig:model}b) the incident right-circularly polarized (RCP) light follows the arc geometry of the right-handed helix and, thus, drives the free electron gas along the curved path.The resulting mode is dipolar (n=1)~\cite{Wozniak2018}. But also for left-circularly polarized (LCP) fields the field vectors may match the local mode pattern for higher field energy and smaller incident wavelengths, respectively. The corresponding mode pattern of the $n=3$ mode can be viewed as line-up of three local dipoles along the helix turn.

This intuitive reasoning can be formalized by analyzing the extinction~\cite{Miller2016}:
\begin{equation}
P_{\textrm{ext},n} = \frac{1}{2} \textrm{Re} \int_V \mathbf{j}_{\text{wire},n}^*\left(\mathbf{r}\right) \cdot \mathbf{E}_\textrm{inc}\left(\mathbf{r}\right) \, \textrm{d}V\, .
\label{eq:scalarprod}
\end{equation}

The total power $P_\textrm{ext}$ transferred from a far-field plane wave $\mathbf{E}_\textrm{inc}$ to any plasmonic scatterer is a direct consequence of Poynting's theorem. It is given by the overlap integral of the incident electric field with the local current density $\mathbf{j}_{\text{wire}, n}$ for each of the excited eigenmodes~\cite{Novotny2012,Miller2016}. The subscript 'ext' refers to 'extinction' as the sum of scattering and absorption. 

Therefore, the overlap integral describes the mode-matching between incident and resonator fields~\cite{Feichtner2017}. According to the identified set of eigenmodes, the current density $\mathbf{j}_{\text{wire},n}$ is given by mapping a sine function with effective wavelength $\lambda_\text{eff}$ onto a helical path. When parameterizing the amplitude of the current along the z direction, the integral \eqref{eq:scalarprod} reduces to a one-dimensional problem (cf. Supplementary Information SIV.). We therefore term our analytical model 1D mode matching. 

The corresponding solution for incident circularly polarized light reads as: 
\begin{align}
&P^\pm_{\textrm{ext},n} = \pm\frac{1}{4} m h E_0 j_{0,n} \times \label{eq:1Dmatching}\\
& \left\lbrace\textrm{sinc}\left[ m h (k \mp k_h - k_\textrm{FP}\right]
- \textrm{sinc}\left[ m h (k \mp k_h + k_{FP}\right]\right\rbrace \nonumber\, ,
\end{align}
where $\textrm{sinc}(x) = \sin(x)/x$. The introduced k-vectors are defined as $k_{FP} = n\pi/mh$, $k_h = 2\pi/h$  and $k = 2\pi/\lambda$ for the excitation wavelength $\lambda$ and the helix pitch height $h$, respectively. $E_0$ and $j_{0,n}$ are the amplitudes of incident field and induced currents for the mode of order $n$. $P^+$ describes excitations with matching handedness, $P^-$ refers to opposite handedness. The subscript of the k-vector $k_\textrm{FP}$ refers to the Fabry-Perot eigenmodes as it can be written according to $k_\textrm{FP} = 2\pi/\lambda^{z}_\textrm{eff}$ with $\lambda^{z}_\textrm{eff}$ being the projection of the effective wavelength onto the helix axis. The relative size of the involved k vectors scales inversely with the respective wavelengths depicted in Figure~\ref{fig:model}~c).

The obtained formula fully quantifies the excitation efficiencies for all types of resonances in a single plasmonic helix. The excitation efficiency of a respective Fabry-Perot eigenmode is determined by its fitting onto the wire length and its match with the exciting external field. Generally spoken, all modes with approximately 1 dipole per turn are efficiently excited by light with the handedness of the helix. All modes with approximately 3 dipoles per turn can be excited by light with the opposite handedness compared to the helix. Beyond the quantitative description, the geometrical considerations of the 1D mode matching provide a straight forward design route for plasmonic helices. 

\begin{figure*}[t]
\centering
\includegraphics[width=\linewidth]{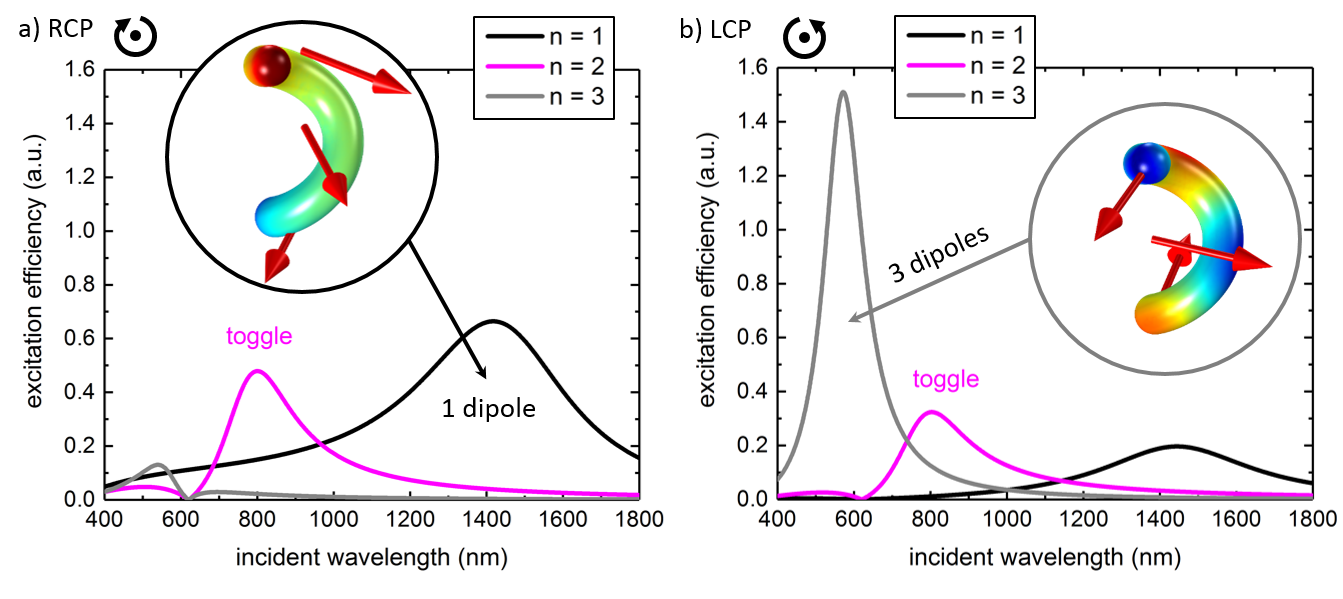}
\caption{\textbf{Excitation efficiencies calculated from the 1D mode matching model for a single-turn helix} (a) Efficient excitation of the fundamental mode occurs for matching handedness of antenna and incident light. (b) For opposite handedness the $n$ = 3 mode can be excited. The insets show numerically calculated plots of surface charge distributions at the surface of the helix and the corresponding incident field vectors. The $n$ = 2 mode is excited for either handedness and marks the toggle point between left-handed and right-handed excitation.}
\label{fig:design}
\end{figure*}

\section{Design Tool}
Based on this we aim to design a plasmonic helix showing both types of resonances, sensitive to matching and opposite handedness, and switching its response in the visible range. We chose silver as the plasmonic metal, a wire radius of around 30\,nm and an excitation wavelength in the middle of the desired range, e.g.~around 800\,nm. According to \eqref{eq:lambdaeff} the corresponding effective wavelength of a linear antenna is 550\,nm. This value will serve as single turn wire length. The wire can be coiled up to a helix with a radius 60\,nm and a single pitch of height 310\,nm. We choose the helix to be right-handed and investigate the obtained resonant behavior using the 1D mode matching model. Figure~\ref{fig:design} shows the excitation efficiencies for the fundamental modes $n = 1 - 3$ dependent on the incident circular polarization state. As the 1D model provides for discrete values of the excitation efficiencies, we plot the corresponding efficiencies by using a Lorentzian 'sample function'. The sample function reflects the match of the respective effective wavelength with the wire length of the 1 turn helix. The insets of Figure~\ref{fig:design} show the results of numerical modeling for the identified resonance conditions of the n=1 and n=3 eigenmodes.

Figure~\ref{fig:design} a) displays the extinction power for RCP light incident on the right-handed helix, i.e. matching handedness. The fundamental dipole mode (n = 1) is most efficiently excited for RCP light at a wavelength around 1450\,nm. This corresponds to an effective wavelength of 1100\,nm doubling the wire length. As can be seen from the inset of Figure \ref{fig:design} a) the incident electric field (red arrows) follows the helical geometry and matches the local charge distribution.  In contrast, the LCP light counteracts the current flow and suppresses the excitation of the $n=1$ mode. The same reasoning applies for the $n = 3$ mode as shown in Figure~\ref{fig:design} b). Here, the LCP excitation drives the charges along the observed mode pattern. Hence, LCP light efficiently excites this mode, while RCP excitation is suppressed.  The external field vectors match with the respective dipole patterns of the numerically modeled surface charge distributions, thereby proving the expected excitation behavior.

The magenta curves in Figure~\ref{fig:design} correspond to the excitation efficiency of the $n = 2$ mode, which deserves special attention. Here, the wire length coincides with the effective wavelength on the wire. In case of the 1-turn-helix this mode can be excited for either handedness of the incident light. The external field drives the charges such that the induced overall dipole moment resembles that of the fundamental resonance of a planar split-ring excited with polarization perpendicular to the gap~\cite{Rockstuhl2006,Wozniak2018}. However, for pitch numbers $m>1$ the dipole moment of one turn may be compensated by the ones of neighboring turns. Hence, light of both handedness' excites this resonance with equal efficiency. As a consequence, this point defines the observed zero crossing of the dissymmetry, and we therefore name it toggle point. 

The identified toggle point can now serve as the starting point for the design of plasmonic helical antennas being resonant in a specific wavelength range for a desired handedness of the incident electric field. One starts by selecting the excitation wavelength where the zero-crossing should occur. Next, the corresponding effective wavelength for the plasmonic wire made of metal a, with thickness b, can be calculated. Finally, this wavelength serves as the single-turn wire length of the envisaged plasmonic helix to be fabricated. Adding further turns to the helix does not change the geometrical excitation conditions. Instead it leads to an increasing number of higher order Fabry-Perot eigenmodes which may be excited according to the reasoning above. 

It has to be noted, that this principle behavior is not dependent on the choice of the metal. Here, we chose silver, since it exhibits the smallest losses of all coinage metals. By switching to e.g.~gold, only the effective wavelength scaling onto the wire has to be adapted to the new material parameters (i.e. the permittivity) to define the wire length. Another point is, that the geometric overlap conditions are not dependent on the radius of the helix. Hence, the radius can serve as an additional degree of freedom and adjusted such that the near-field interaction between neighboring helix turns is negligible but the extinction power is maximized.

\begin{figure*}[t]
\centering
\includegraphics[width=0.7\linewidth]{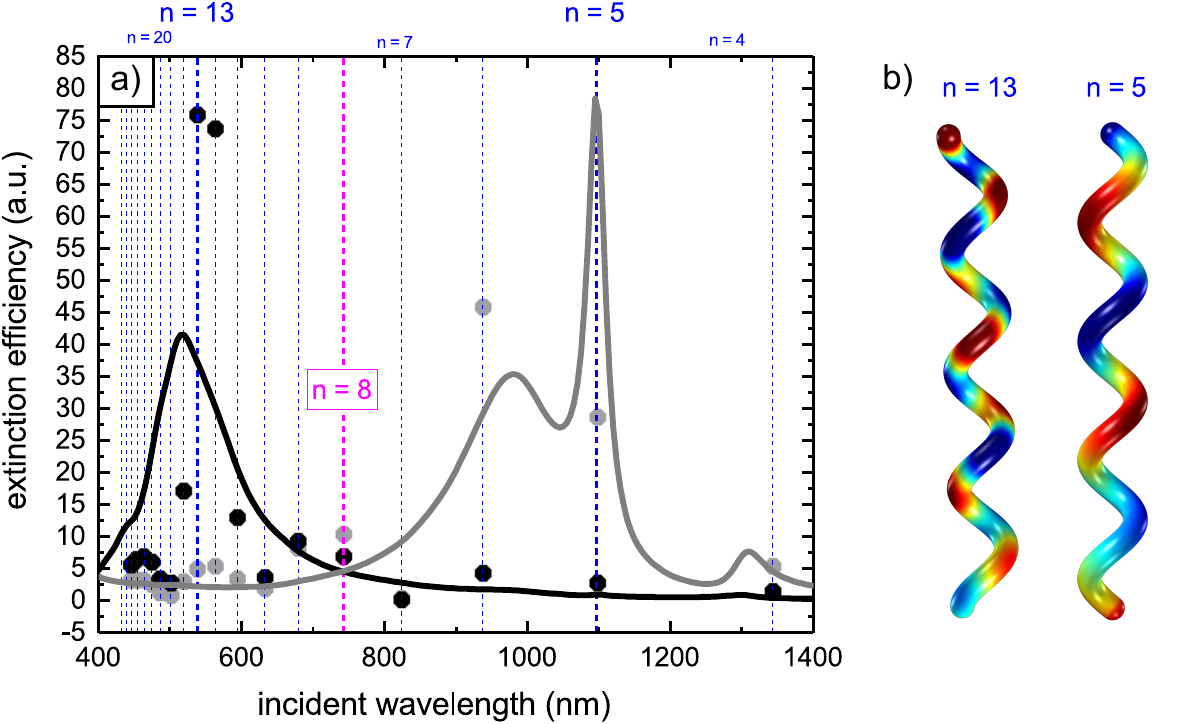}
\caption{\textbf{Comparison of 1D mode matching to full-field modeling} a) Analytically determined resonance positions and strengths  for a right-handed 4 turn helix are displayed as dots. The dot height is proportional to the area of the excited modes. The curves display the corresponding extinction efficiencies from full-field modeling. For both, dots and curves, black (gray) corresponds to incidence of left-(right-)circularly polarized light.  (b) The plotted surface charges distributions from modeling show the expected mode order.}
\label{fig:comparisonFEM}
\end{figure*}

Based on the identified geometrical parameters, Figure~\ref{fig:comparisonFEM} compares the analytical model to full-field simulations for a helix with 4 turns ($m = 4$).
The dashed blue lines in Fig.~\ref{fig:comparisonFEM}a) display analytically calculated Fabry-Perot modes of a straight silver cylinder with a radius of 32\,nm, a length of 2016\,nm and round end caps. They agree well with the respective modes retrieved from full field modeling (not displayed in this graph).
The dots in Figure~\ref{fig:comparisonFEM}~a) display the efficiency of energy transfer to the respective Fabry-Perot eigenmodes dependent on the handedness of the incident light according to equation~\eqref{eq:1Dmatching}. Black dots stand for LCP excitation, grey ones for RPC excitation. The mode matching has been performed for the exact resonance positions for sinusoidal helix currents. The quantitative agreement only requires one additional parameter. The helix radius was corrected according to $r + \delta r$ to account for the near-surface character of plasmonic currents for higher mode orders; $\delta r$ was chosen to be 6\,nm (see Supplementary Material SIV). The numerically calculated extinction efficiencies $Q_\text{ext}$ of a right-handed helix with 4 turns are plotted as black (LCP) and gray (RCP) curves Fig.~\ref{fig:comparisonFEM}~a). The chiroptical response is characterized by efficient excitation of resonant modes around 1000\,nm for RCP light, a transition region with no excitations around 800\,nm and the LCP excitation range around 600\,nm. Both ranges of strong dissymmetry in Fig.~\ref{fig:comparisonFEM}~a) contain plasmonic Fabry-Perot modes of different orders as can be seen in Fig.~\ref{fig:comparisonFEM}~b), showing surface charge distributions for two selected modes. At $\lambda=1070\,$nm RCP light excites a mode of order $n = 5$, while for LCP excitation at $\lambda=530\,$nm the excited mode has the order $n= 13$. The 1D design tool provides the envelope of geometrically excitable modes and determines the excitation efficiencies of the underlying Fabry-Perot eigenmodes (see Supplementary Information SV).

\section{Experimental realization}
\begin{figure*}[tb]
\centering
\includegraphics[width=\linewidth]{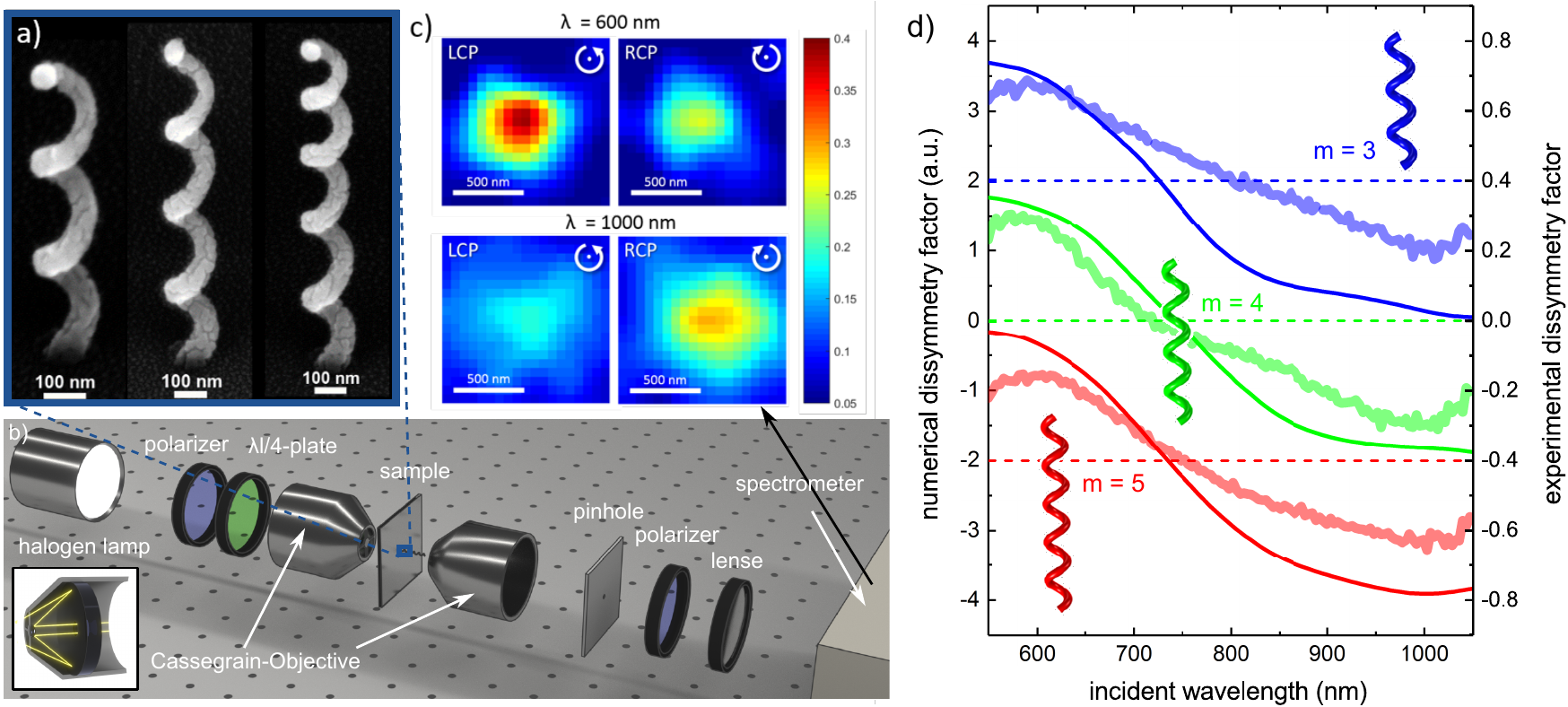}
\caption{ \textbf{Single plasmonic helices in experiment} a) Scanning electron micrographs show prototypical helical plasmonic nanoantennas. The investigated silver helices have 3- 5 turns for a radius of 60\,nm and a wire radius of 32\,nm. b) Circularly polarized light is focused onto the sample by a Cassegrain objective. A second Cassegrain focuses the transmitted light onto a pinhole which cuts away the out-of-focus signal contribution to allow for spectroscopic investigation of single nanostructures. c) Extinction maps of a helix with 4 turns for two different wavelengths under incidence of both circular polarization states. While at 1000\,nm the right-handed helix responds to right-circularly polarized (RCP) light, the helix is sensitive to left-circularly polarized light (LCP) at 600\,nm. d) Comparison of measured dissymmetry factors to full field modeling.}
\label{fig:exp}
\end{figure*}
Finally, the designed helical plasmonic geometry is realized experimentally. Figure~\ref{fig:exp}~a) shows scanning electron micrographs of the fabricated right-handed silver helices with $m = 3-5$ turns, with a helix radius of $r_h=60\,$nm and a wire radius of $r_w=32\,$nm, as determined from the SEM images. The employed setup for confocal white light transmission spectroscopy is shown in Figure~\ref{fig:exp}~b). Two Cassegrain-objectives allow for dispersion free extinction spectroscopy of single nanostructures in the wavelength range from 550$\,$nm to 1050$\,$nm. The helices were directly written onto transparent substrates using electron beam induced deposition (EBID), currently being the most flexible and precise 3D nanostructuring technique\cite{Utke2008,Hoeflich2011, Haverkamp2017, Fowlkes2016}. Silver was subsequently deposited onto the EBID scaffold by means of glancing angle deposition~\cite{Wozniak2018,Hoeflich2012}. The thickness of the resulting silver shell exceeds the skin depth, such that an optical description as solid silver helix is valid. 120 helices with different numbers of turns were spectroscopically examined for incident left- and right-circularly polarized (LCP and RCP) broadband light in the range of $\lambda =$ 500 -- 1100\,nm (see Supplementary Information SII). 

For the case of a 4 pitch helix Figure~\ref{fig:exp}~c) displays extinction maps for both incident polarizations at two selected wavelengths of $\lambda = 600\,$nm and $\lambda = 1000\,$nm. According to the design a strong response of the right-handed helix to RCP light at 1000\,nm occurs, while the response to LCP light is maximized at 600 nm. We quantify the strength of the chiroptical response by the dissymmetry factor $g$. Physically, $g$ describes the difference in the extinction for different handedness of incident circularly polarized light normalized by the total extinction:
\begin{equation}
g_\text{Cext}  = \frac{2 \left(C_{\text{ext}}^{\text{LCP}}-C_{\text{ext}}^{\text{RCP}}\right)}{C_{\text{ext}}^{\text{LCP}}+C_{\text{ext}}^{\text{RCP}}}
\label{eq:g_Cext}
\end{equation}
with the calculated extinction cross sections $C_{\text{ext}}^{\text{LCP}}$ and $C_{\text{ext}}^{\text{RCP}}$  from the scattering problem~\cite{Bohren}. This directly translates to:
\begin{equation}
g_\text{1-T}= \frac{2 \left(T_{\text{RCP}}-T_{\text{LCP}}\right)}{2-T_{\text{LCP}}-T_{\text{RCP}}} ,
\label{eq:g_1-T}
\end{equation}
for experimentally measured transmission spectra $T_{\text{LCP}}$ and $T_{\text{RCP}}$. Since both definitions are equivalent ($g_\text{Cext} = g_\text{1-T} = g$), equations \eqref{eq:g_Cext} and \eqref{eq:g_1-T} provide a quantitative comparison between theoretically expected and experimentally realized chiroptical response (cf. Supplementary Information SI.). Furthermore, the extremal values of $g_\text{Cext}$ coincide with the extremal values of the measure of electromagnetic (EM) chirality~\cite{Fernandez-Corbaton2016} and hence can be maximized in terms of chiroptical response. 

The obtained dissymmetry factors of the investigated helices with $m = $ 3, 4 and 5 turns are plotted in Fig.~\ref{fig:exp}~d). Solid lines display the average experimental dissymmetry factors $g_\text{1-T}$ of 30 single helix measurements for each number of turns. The dashed lines display $g_\text{Cext}$ and are calculated based on full field modeling with the geometrical parameters taken from SE micrographs. The modeling reproduces the observed spectral behavior well. The maximum value of the dissymmetry factor deduced from the simulations is close to the theoretical limit of 2. Experimentally, much smaller dissymmetry factors around 0.45 are observed. This discrepancy is mainly due to the chosen experimental setting in which light is focused to a spot size that exceeds the helix diameter (cf. Supplementary Material SII). Furthermore,  scattering losses due to surface roughness and grain boundaries may cause an additional achiral loss background and thereby decrease the experimentally observed chiroptical response. The small spectral deviations can be attributed to geometrical imperfections of the real structures which have a larger influence for the single helix measurement. Despite these limitations, the experimentally measured dissymmetry is the highest observed up to now for a single plasmonic nanohelix (see Supplementary Information SII). 

\section{conclusions}
Here, the interaction of single plasmonic helices with circularly polarized light was investigated, both theoretically and experimentally.  Our findings lead to a complete understanding of all resonant features of plasmonic helices of any geometry and material. An analytical 1D mode matching model was developed that quantitatively describes the excitation efficiencies of all types of resonant modes. Furthermore, the mode matching provides a straightforward design route for plasmonic helices being resonant in a desired wavelength range. We have employed the developed design tool to design helical plasmonic antennas with a strong chiroptical response for either handedness of incident light in the visible range. The obtained helical geometry was experimentally realized and showed an extraordinary high dissymmetry for resonant modes responding to both handedness of incident circular polarization. In conclusion, an efficient route for the design and fabrication of helical nanoantennas is established.  Thereby, potentially groundbreaking applications of plasmonic helices in all-optical information processing may be triggered, either based on the spin angular momentum of light~\cite{Krauss} or based on their strong nonlinear response~\cite{Rodrigues2017}.

\section{Methods}
Glass cover slips were sputter-coated with a 50~nm thick amorphous layer of indium tin oxide (AJA International INC.) to serve as transparent conductive substrates for direct electron beam writing. Helices were written in a FEI Strata DB 235 dual beam focused ion beam microscope using the metal-organic precursor dimethyl-gold-acetylacetonate (Me$_2$Au(acac)) by directly addressing the patterning board using streamfiles. Therewith, the electron beam was scanned in a circular path with a radius of 60\,nm, a pitch of 600~pm and a stepwise increasing dwell time $t_d = $ 8~ms $+ (n - 1) \times $ 4~ms, with the number of turns $n$. The applied step function compensates for the decrease of the vertical growth rate in the deposition with increasing structure height. The resulting helices have a constant pitch height of around 310~nm and arm thicknesses slightly below 20~nm and were used as a scaffold to be subsequently covered with the desired metal. Conformal coverage with silver was achieved by sputter coating (Plassys) under glancing angle incidence with and emission current of 110~mA and a deposition rate of $0.4~$nm/s. The evaporated film thickness of 50~nm resulted in a silver layer of $~$21.5~nm onto the helices. This shell thickness exceeds the skin depths making their plasmonic response in the visible range equivalent to that of a pure silver helix~\cite{Haverkamp2017}. The resulting arm thickness of the helices is 64\,nm.

Optical spectroscopy of single silver helices was carried out using a home-built confocal microscope. The collimated  and IR-filtered emission from an incoherent halogen lamp covering the spectral range from 400\,nm to 1000\,nm (Dolan-Jenner Model DC-950) is focused onto the sample  using an all-reflective Cassegrain objective (ARO) with a numerical aperture of 0.5 (5002-000, Beck Optronics Solutions Ltd.) resulting in a spot with a diameter of approximately 50\,$\upmu$m. Left- and right-circularly polarized light in the complete visible spectral region was generated by a combination of a Glan-Taylor linear polarizer (PGL 15, B. Halle Nachf.) and a rotatable superachromatic quarter wave plate (WPA4415, Union Optics).
The light transmitted through the sample is collected by a second ARO (5006-000, Beck Optronics Solutions Ltd.) and imaged onto a pinhole with a diameter of 75\,$\upmu$m in the backfocal plane of the ARO, limiting the observation area on the sample to approximately 1\,$\upmu$m. A 100\,mm lens is afterwards used to focus the light onto the entrance slit of a spectrometer (SP2150, Princeton Instruments) in combination with a liquid-nitrogen cooled CCD camera (Pixis eXcelonBR, Princeton Instruments).
By raster-scanning the sample with the helices through the focus with a nominal step size of 250\,nm perpendicular to the propagation direction of the beam using a 3D piezo scanner (P-611.3 Nanocube von Physik Instrumente) maps of the linear extinction of the silver helices are obtained.

Full field modeling was carried out based on the finite element method (Comsol Multiphysics) and on the finite difference time domain technique (Lumerical Solutions) with the material response taken from reference~\cite{Johnson1972}. The corresponding models were optimized concerning accuracy and convergence, the calculated spectra agreed perfectly for both techniques (see Supplementary Information SV). The helices were modeled with round end caps and geometrical parameters taken from the SEM images. Analytical calculations were carried out using Mathematica and later transferred to Python. The linear wavelength scaling of Novotny~\cite{Novotny2007} has been included in the corresponding code. The electromagnetic near-fields of different helix turns may interact and shift the excitation wavelength compared to the linear cylinder. Therefore, the near-field extension around the wire was calculated to define a minimum pitch height for the helix down to which the analytical treatment is justified (see Supplementary Information SIII). The developed analytical design tool allows for identifying an appropriate helix geometry for a desired resonant wavelength range and is available for download~\footnote{Python scripts available: \tt{https://sourceforge.net/} \tt{projects/plasmonic-helix-1dmodel/}}. 

\section{Funding Information}

KH acknowledges funding from the Helmholtz association by the Postdoctoral Fellowship PD-140, MS acknowledges funding from the Federal Ministry of Education and Research within the NanomatFutur program. Furthermore, financial support by the Deutsche Forschungsgemeinschaft (SPP1839 “Tailored Disorder”) and the German-Israeli Foundation (GIF grant no. 1256) is gratefully acknowledged.

\section{Acknowledgments}
The authors thank Pawel Wozniak and Ivan Fernandez-Corbaton for fruitful discussions.
\section{Author contributions}
KH initiated the study, developed and supervised the helix fabrication, and wrote manuscript and SI. TF conceived the mode matching. TF and KH performed the analytical and numerical calculations. EH, HK, MS and CL developed the optical measurement scheme, performed the measurements, and evaluated the data. CH fabricated the transparent substrates and helices. All authors discussed the results and contributed to finalizing the manuscript.

\bibliography{silverHelix}

\end{document}



\title{\Large{Supplementary Information:}\\[4mm]\large{The resonant behavior of a single plasmonic helix}}

\author{K. H\"{o}flich}
\email[Corresponding author: ]{katja.hoeflich@helmholtz-berlin.de}
\affiliation{%
Helmholtz Zentrum f\"{u}r Materialien und Energie Berlin,\\ Hahn-Meitner-Platz 1, D -- 14109 Berlin, Germany
}%
 \affiliation{Max Planck Institute for the Science of Light}
\author{Th. Feichtner}%
 \affiliation{%
Nano-Optics \& Biophotonics Group, Department of Experimental Physics 5, R\"{o}ntgen Research Center for Complex Material Research (RCCM), Physics Institute, University of W\"{u}rzburg, Am Hubland, D-97074 W\"{u}rzburg, Germany
}%
\author{E. Hansj\"{u}rgen}%
\affiliation{%
 Institute of Physics and Center of Interface Science, Carl von Ossietzky Universit\"{a}t Oldenburg, D-26129 Oldenburg, Germany
}%
\author{C. Haverkamp}%
\affiliation{%
Helmholtz Zentrum f\"{u}r Materialien und Energie Berlin,\\ Hahn-Meitner-Platz 1, D -- 14109 Berlin, Germany
}%
 \affiliation{Max Planck Institute for the Science of Light}

\author{H. Kollmann}
\author{C. Lienau}
\author{M. Silies}
\affiliation{%
Institute of Physics and Center of Interface Science, Carl von Ossietzky Universit\"{a}t Oldenburg, D-26129 Oldenburg, Germany
}%

\date{\today}
\begin{abstract}
Chiral plasmonic nanostructures will be of increasing importance for future applications in the field of nano optics and metamaterials. Their sensitivity to incident circularly polarized light in combination with the ability of extreme electromagnetic field localization renders them ideal candidates for chiral sensing and for all-optical information processing either based on the spin angular momentum of lightor based on its strong nonlinear response. Here, the resonant modes of single plasmonic helices are investigated. We find that a single plasmonic helix can be efficiently excited with circularly polarized light of both equal and opposite handedness relative to the handedness of the helix. An analytic model provides resonance conditions matching the results of full-field modeling. The underlying geometric considerations explain the mechanism of excitation and deliver quantitative design rules for plasmonic helices being resonant in a desired wavelength range. Based on the developed analytical design tool single silver helices were fabricated and optically characterized. They show the expected strong chiroptical response to both handednesses in the targeted visible range. The thereby experimentally realized dissymmetry factor is the largest obtained for single plasmonic helices in the visible range up to now. 
\end{abstract}             
\maketitle

\renewcommand{\theequation}{S\arabic{equation}}
\renewcommand{\thesection}{S\Roman{section}}
\renewcommand{\thefigure}{S\arabic{figure}}

\section{Quantification of chiroptical response}
\label{chiroquant}
The strength of the chiroptical response can be quantified in different ways. Optical rotation and ellipticity are typically defined by the polarization ellipse of the transmitted light obtained upon incidence of linearly polarized light~\cite{Barron:2009:MLS:1816638}. While optical rotation refers to the rotation of the polarization plane, ellipticity is given by the ratio of the major axes of the resulting polarization ellipse. Since both quantities are Kramers-Kronig related, the opto-chiral interaction can be quantified using either of them. The ellipticity can be determined by the difference of the amplitudes of transmitted circularly polarized light according to:
\begin{equation}
\text{tan}\psi = \left(|\mathbf{E}|_{\text{LCP}} - |\mathbf{E}|_{\text{RCP}}\right)/ \left(|\mathbf{E}|_{\text{LCP}} + |\mathbf{E}|_{\text{RCP}}\right).
\label{eq:tanpsi}
\end{equation}
Often the ellipticity $\psi$ is measured in terms of the square roots of the corresponding transmitted intensities of LCP and RCP incident light. It has to be mentioned that this is valid only if the conversion of LCP into RCP and vice versa is negligible. Especially, for the case of plasmonic helices this is not necessarily true~\cite{Gansel2010}. The ellipticity is related to the circular dichroism, which refers to the difference in the extinction coefficients $\kappa_{\text{LCP}}$  and $\kappa_{\text{RCP}}$  for incident LCP and RCP light. Here, special care has to be taken concerning the terminology. In chemistry often the term absorption coefficient is used for $\kappa$. In the same manner absorbance is used in chemical literature when actually the extinction (i.e. the sum of absorption and scattering) is measured. The typical experimental measure is the transmittance under LCP and RCP incidence. The corresponding definition in chemistry is the molar extinction coefficient
\begin{equation}
\epsilon^\text{chem} = -\frac{1}{cl}\text{ln} T
\label{eq:extchem}
\end{equation}
which is normalized to the molar concentration of molecules per liter $c$ and the optical path length $l$. In case of large ensembles of plasmonic objects the use this quantity and the resulting relationship to the ellipticity ($\psi\left[\text{mdeg}\right] = 3300 \left(\epsilon_\text{LCP}-\epsilon_\text{RCP}\right)$) may be justified in terms of an effective medium. Still, proper normalization is an issue and may lead to arbitrarily high values of ellipticity when omitted~\cite{Yin2013}. 
To obtain a reliable comparison between the opto-chiral activity of both chiral matter as well as single chiral objects the dissymmetry factor $g$ can be used~\cite{Kuhn1930}. The dimensionless quantity
\begin{equation}
g = \frac{2\left(\epsilon_{\text{LCP}}-\epsilon_{\text{RCP}}\right)}{\epsilon_{\text{LCP}}+\epsilon_{\text{RCP}}}.
\label{eq:g}
\end{equation}
is normalized by the total extinction. Since $g$ does not depend on the specific way of normalization of the extinction, it is easily transferable to either extinction coefficient with $\epsilon \propto -\text{ln}T $. 

To compare experimentally obtained dissymmetries to the corresponding theoretical values, special care has to be taken. For single scatterers the extinction cross section $C_\text{ext}$ is defined by integrating the radial projection of the energy transfer rate over the spherical surface A in the far-field and normalized by the incident irradiance (intensity) $I_\text{inc}$~\cite{Bohren}:
\begin{equation}
C_\text{ext} = C_\text{abs} + C_\text{sca} = -\frac{1}{I_\text{inc}} \int_\text{A}\mathbf{S}\cdot\mathbf{e}_r dA + \frac{1}{I_\text{inc}}\int_\text{A}\mathbf{S}_\text{sca}\cdot\mathbf{e}_r dA
\label{eq:extBohren}
\end{equation}
The Poynting vectors $\mathbf{S}$ and $\mathbf{S}_\text{sca}$ describe the energy transfer rates of the total and scattered electromagnetic fields. According to the optical theorem the extinction cross section depends on the scattering amplitude in forward direction only and, hence, determines the decrease of the detector area $A(D)$ by the object due to scattering and absorption in transmission measurements:
\begin{equation}
T^\text{exp} = \frac{I_\text{T}}{I_\text{inc}} = \frac{A(D)-C_\text{ext}}{A(D)}
\label{eq:Texp}
\end{equation}
For small values of $C_\text{ext}$ compared to the detector area, this coincides with the negative decadic logarithm typically taken in optical measurements:
\begin{equation}
C_\text{ext}/A(D) \approx \epsilon^\text{exp} = - \text{ln} T^\text{exp}
\label{eq:C/A}
\end{equation}
In our measurement setup the (virtual) detector area is given by the circular area of the pin hole in the image plane (since this area fills only part of the physical detector area). With 1.63\,$\upmu$m$^2$ the detector area has almost the same order of magnitude as the measured cross-sections. Hence, the logarithm enlarges $\epsilon^\text{exp}$ relative to $C_\text{ext}/A(D)$. The influence on the dissymmetry factors is minor in our case. Nevertheless, in terms of quantitative comparison of the experiment with numerically or analytically obtained values, we employ the most appropriate choice for the dissymmetry factor given by:
\begin{equation}
g_\text{Cext} = g_\text{1-T} = \frac{2 \left(C_{\text{ext}}^{\text{LCP}}-C_{\text{ext}}^{\text{RCP}}\right)}{C_{\text{ext}}^{\text{LCP}}+C_{\text{ext}}^{\text{RCP}}}= \frac{2 \left(T_{\text{RCP}}-T_{\text{LCP}}\right)}{2-T_{\text{LCP}}-T_{\text{RCP}}}.
\label{eq:g1-T}
\end{equation}
Here again, the normalization is carried out with respect to the total extinction. Furthermore, the detector area cancels out making the quantity independent from a specific measurement scheme. Thereby, $g_\text{1-T}$ is suitable not only for single particle measurements but also for measurements on ensembles of chiral objects. Physically, both of the dissymmetry factors $g$ and $g_\text{1-T}$ account for the fact that the same absolute difference in extinction can be caused by a strong or weak chiroptical response.

In terms of optimizing the opto-chiral interaction, the definition of EM-chirality is very interesting, since it can be sorted in ascending order and provides an upper limit~\cite{Fernandez-Corbaton2016}. An object is maximally EM chiral if and only if it is transparent to light of one handedness and does not change the helicity of the fields upon interaction~\cite{Fernandez-Corbaton2016}. This mathematical description nicely resembles our choice for the dissymmetry factor $g_\text{1-T}$ which achieves its upper bound $\pm 2$ if $T_\text{LCP} = 1$ and $T_\text{RCP} = 0$ or vice versa. The chemical definition of $g$ is mathematically analogue (since the contained logarithm is monotone) and leads to a value of 0.9 reflecting the choice of the exponential. 

Here, it has to be mentioned that employing the transmission to calculate the dissymmetry factor according to~\cite{Esposito2014,Kosters2017} 
\begin{equation}
g_\text{T} = \frac{2 \left(T_{\text{RCP}}-T_{\text{LCP}}\right)}{T_{\text{LCP}}+T_{\text{RCP}}}.
\label{eq:gT}
\end{equation}
contradicts its physical meaning. A generally strongly absorbing material with a small difference in transmission for LCP and RCP incidence provides larger values of $g_\text{T}$ than a material with the same asymmetry ($T_{\text{LCP}}-T_{\text{RCP}}$) but weaker polarization independent absorption.
As a consequence, the definition $g_\text{T}$ achieves its maximum value of 2 not only for maximally EM chiral objects but also for objects with a strong achiral absorption background and/or polarization conversion (e.g. $T_\text{LCP} = 0.1$ and $T_\text{RCP} = 0$).

\begin{table}
\definecolor{lightgray}{gray}{0.9}
\setlength{\extrarowheight}{4pt} 
\setlength{\tabcolsep}{5pt}
\newcolumntype{g}{>{\columncolor{lightgray}}c}
\centering
\begin{tabular}{ |l | l  l |g|  l| }
  \colrule	
  $\lambda [\text{nm}] $&$g_\text{T}$ & $g$&$g_\text{1-T}$ & reference\\
   \colrule	
  - & 2 & 0.92& 2&maximally EM chiral object ($T_\text{LCP}=0$, $T_\text{RCP}=1$)\\
  - & 2 & 0.01& 0.01&object with strong achiral background($T_\text{LCP}=0$, $T_\text{RCP}=0.01$)\\
    \colrule	
  4300 & -1.27 & -0.67& -1.56&\cite{Thiel2010}, ensemble of photonic helices from photoresist, not transferable to VIS\\
  3500 & 1.66 & 0.74& 1.47&\cite{Gansel2009}, ensemble of left-handed gold helices in low loss IR range\\
  600 & 0.14 & 0.10& 0.33&\cite{Esposito2014}, ensemble of helices from carbon platinum mixture, electron beam writing\\
  640 & 0.22 & 0.05& 0.06&\cite{Esposito2014}, ensemble of helices from carbon platinum mixture, ion beam writing\\
  995 & -0.41 & -0.14& -0.21&\cite{Esposito2014}, ensemble of helices from carbon platinum mixture, ion beam writing\\
  700 & 0.73 & 0.23& 0.34&\cite{Esposito2015}, ensemble of triple helices, platinum carbon mixture, ion beam writing\\
   530 & -0.67 & -0.11& -0.13&\cite{Esposito2016}, ensemble of helices from carbon platinum mixture, ion beam writing\\
  840 & 0.27 & 0.07& 0.09&\cite{Esposito2016}, ensemble of helices from carbon platinum mixture, ion beam writing\\
  495 & -0.41 & -0.12& -0.18&\cite{Esposito2016}, ensemble of helices from carbon, ion beam writing\\
  750 & 0.5 & 0.24& 0.46&\cite{Esposito2016}, ensemble of helices from carbon, ion beam writing\\
      \colrule	
    550 & - & - & 0.45&this work, silver-coated single helices from electron beam writing\\
  1000 & - & - & 0.30&this work, silver-coated single helices from electron beam writing\\
  1450&-0.03 &-0.028 & -0.43& \cite{Wozniak2018}, gold-coated single helix with 1 turn in near IR\\
   \colrule
    520 & -0.52 & -0.19& -0.30&\cite{Esposito2016}, modeled ensemble of helices from carbon\\
  800 & 0.45 & 0.23& 0.47&\cite{Esposito2016}, modeled ensemble of ensemble of helices from carbon\\
       850 & - &-& 0.8&\cite{Singh2018}, modeled core-shell helix ensemble with a-Si shell (transmission spectra not provided)\\
    695 & -1.13 & -0.51& -0.96&\cite{Kosters2017}, modeled left-handed gold-coated helix array (normalization issues in experiment)\\
   1000 & 1.64 & 0.70& 1.32&\cite{Kosters2017}, modeled left-handed gold-coated helix array (normalization issues in experiment)\\
     1450&-0.03 &0.028 & 0.5& \cite{Wozniak2018}, modeled single gold helix of 1 turn\\
      550 & - & -& 1.85& this work, modeled single silver helices\\
  1050 & - & -&1.91& this work, modeled single silver helices\\
  \colrule	
\end{tabular}
\caption{\label{tab:1}\textbf{ Overview of the achieved values for different definitions of the dissymmetry factors} The top rows show theoretically achievable values, followed by experimental investigations of ensembles of helices. The next rows display experimental values of single plasmonic helices, followed by numerically calculated values. The values from external references were calculated from printed transmission spectra. }
\end{table}

Table \ref{tab:1} shows a comparison of all three dissymmetry factors calculated from the literature values of transmission. The largest dissymmetry values $g_\text{1-T}$ for plasmonic systems could be achieved  in the near- and infrared region, where metallic losses are negligible\cite{Gansel2009,Yin2013}. Dielectric chiral metamaterials showed even larger dissymmetry factors\cite{Thiel2010} but are typically limited to the infrared region. In the visible range silver helices are the near optimum choice providing dissymmetry factors close to 2 in full field simulations with and without substrate (ITO-coated glass). For the given geometrical parameters a single silver helix of 5 turns (without substrate) achieves a maximum dissymmetry of around 1.85 for LCP light of 550\,nm and of 1.91 for RCP light at 1000\,nm. Interestingly, carbon helices showed remarkable dissymmetry factors as well which deserve further investigation. However, it has to mentioned that these helices already work close to their optimum performance as can be seen from Table \ref{tab:1}. The same holds true for the investigated single turn gold helices. The achieved dissymmetry factor in the near-IR is remarkable but already close to the numerically predicted value\cite{Wozniak2018}. Hence, exploiting the full chiroptical potential of silver helices is of great scientific interest for applications in chiral sensing and all-optical information processing.

\section{Evaluation of transmission measurements}
\begin{figure}[t!]
\centering
\includegraphics[width=0.7\linewidth]{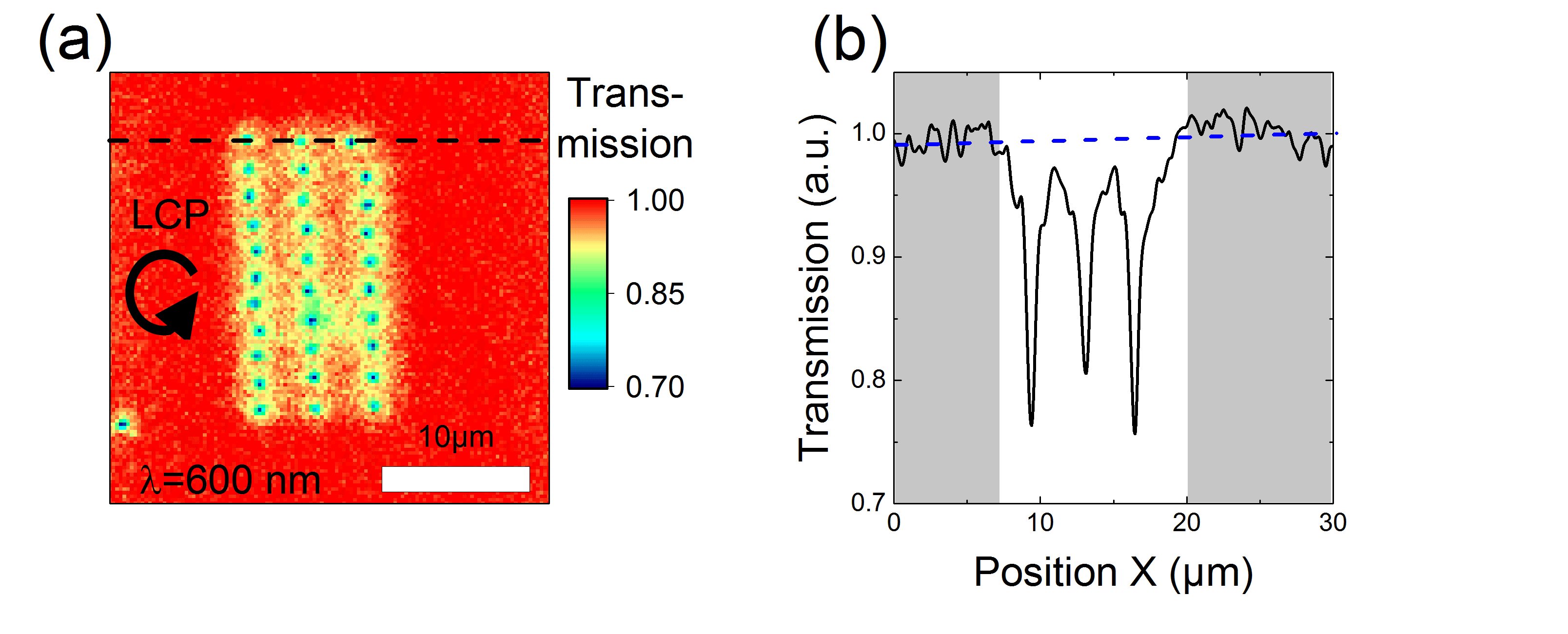}
\caption{\textbf{Normalization scheme of optical transmission measurements} (a) Transmission map of an array of 30 helices with 4 pitches at a wavelength of 600\,nm for excitation with left-circularly polarized light. The helices are visible by the lowered transmission. 
(b) Line scan along the black-dashed line in (a) with three helix transmission dips down to 0.75. The spatially-dependent background along the x-axis, indicated by the blue-dashed line is determined by taking into account the grey-shaded regions only.
}
\label{fig:expnorm}
\end{figure}

Two-dimensional extinction spectra with a spatial step size of 250\,nm and a spectral resolution of 3.7\,nm are obtained by raster-scanning the sample with the silver helices through the focus of an all-reflective objective (ARO), resulting in three-dimensional data sets for left and right-circular polarized light. The measurement procedure for each polarization state takes approximately 30 minutes. In Fig.~\ref{fig:expnorm} (a), a prototypical transmission map of the array is shown at a wavelength of 775\,nm for right-circular polarized light. The spatial resolution of the confocal microscope is improved to approximately 1 µm by placing a 75 µm pinhole in the backfocal plane of the ARO. Afterwards, a position-dependent background signal is subtracted for each of the 137 wavelength intervals that is caused by slight irregularities on the sample substrate, as can be seen in Fig. \ref{fig:expnorm}(b) for the line scan shown in (a). The resulting two-dimensional maps for each wavelength are afterwards normalized to the reference transmission spectrum collected using a cover glass as the sample.

\begin{figure}[b]
\centering
\includegraphics[width=0.85\linewidth]{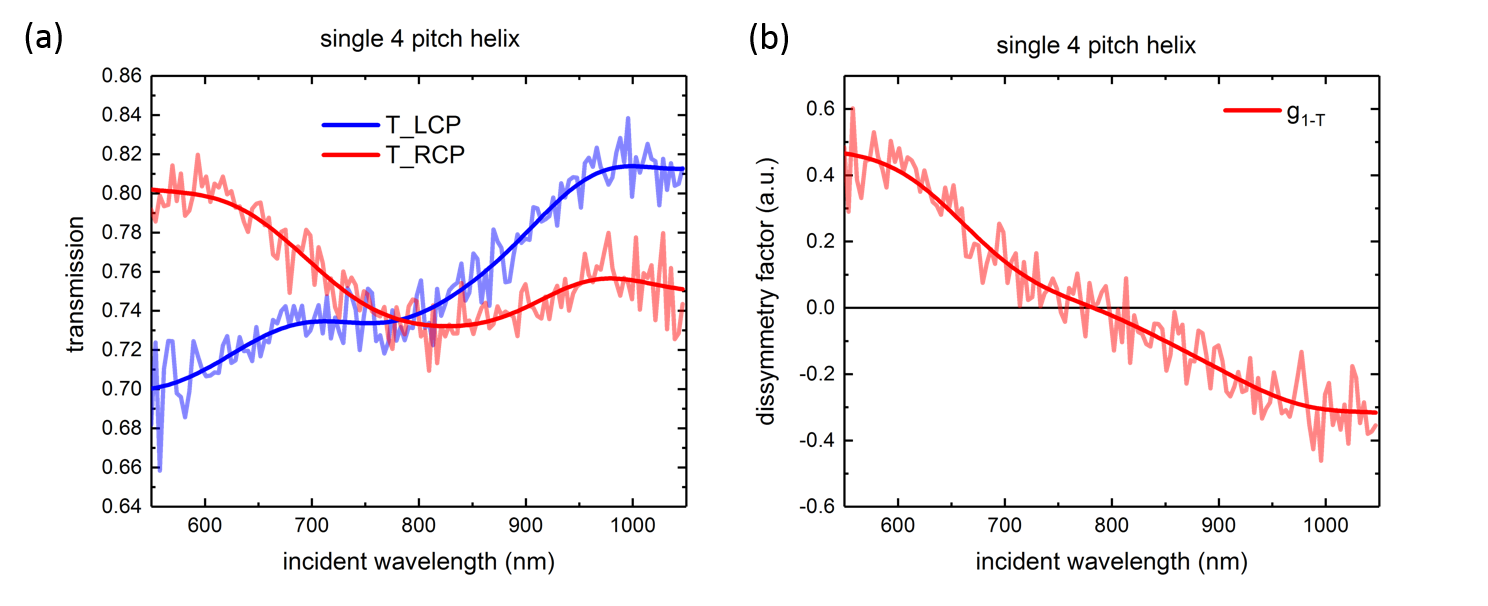}
\caption{\textbf{Single helix measurement} a) Transmission measurement of a 4 turn helix and b) resulting dissymmetry factor. The smoothed curves serve as guide for the eye.}
\label{fig:h23}
\end{figure}

In order to quantify the experimental dissymmetry factor g, the total accumulated transmission of each silver helix is essential. However, small spatial fluctuations of the sample during the scanning procedure resulted in slight, but non-negligible variations of the spatial position of the transmission minimum of the silver helices for left- and right-circular polarized light. Hence, the total transmission for each wavelength interval is determined by spatially integrating the transmission in a small region around the transmission minimum and the dissymmetry factor is calculated for every wavelength using equation \eqref{eq:g1-T}. Now, transmission spectra for each silver helix and the respective handedness of incident light are obtained by recomposing the values for each wavelength. Figure \ref{fig:h23} shows the transmission spectra (a) and the dissymmetry factor (b) of a prototypical 4 pitch helix. Since here we present single helix measurements the signal to noise ratio is poor. Hence, for the experimental dissymmetries in the main manuscript, all single helix measurements were taken (30 of each geometry) and averaged to avoid any biased selection.

\newpage

\section{Linear wavelength scaling}
The calculated eigenmodes of the straight cylindrical wires are justified as helix eigenmodes for negligible interaction of the different helix turns. In other words, if $h>>d_{\text{near-field}}$ defining the distance in which the near-fields decayed by $1/e$. In case of a loss-less Drude model for the permittivity of silver (according to Novotny) with $\epsilon_\infty = 4.0$ and $\omega_p=9.16 \text{eV}$ this requires a minimum pitch height of 200\,nm for the visible range.

\begin{figure}[t]
\centering
\includegraphics[width=0.85\linewidth]{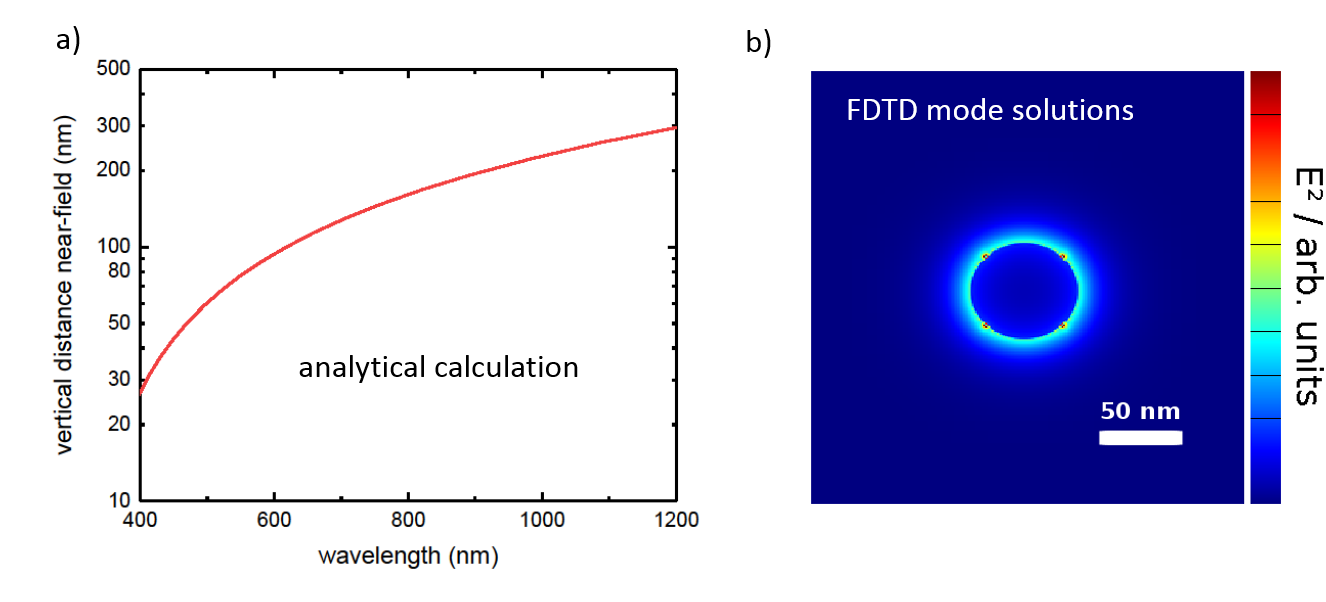}
\caption{\textbf{Decay length of the plasmonic near-field perpendicular to the antenna surface} a) The decay length plotted is the distance at which the near-field has decayed by 1/e. b) Plot of the absolute values of the electric nearfield of the fundamental mode in a infinitely long straight silver wire of radius 32\,nm at a wavelength of $\lambda = 400$\,nm.}
\label{fig:dnear}
\end{figure}

Figure \ref{fig:dnear} shows the wavelengths dependent radial decay of the near field around a cylindrical silver wire of 32\,nm radius. The fields are decayed exponentially to $1/e$ at radial distances $\lesssim$ 200\,nm throughout the visible range. To ensure a negligible contribution of the interaction between the different helix turns, the pitch heights of the fabricated helices were chosen  to be around 300\,nm. The linear wavelength scaling formalism for short plasmonic wires of Novotny~\cite{Novotny2007} is used within this work, more details can be found in Ref. \cite{Su2015}.

\section{1D-model}

The calculation is performed in the frequency domain, w.l.o.g. assuming time-harmonic fields. By using the phasor formalism, the $e^{-i\omega t}$ terms are suppressed. Poynting's theorem can be used to describe the extinction power $P_\text{ext}$ when far-field radiation $\mathbf{E}_\text{inc}$ excites currents $\mathbf{j}$ in a resonator or an optical antenna~\cite{Novotny2012}:
\begin{equation}
P_{\textrm{ext},n} = \frac{1}{2} \textrm{Re} \int_V \mathbf{j}^*\left(\mathbf{r}\right) \cdot \mathbf{E}_\textrm{inc}\left(\mathbf{r}\right) \, \textrm{d}V\,
\label{eq:scalarprod2} \, ,
\end{equation}
where $^*$ denotes the complex conjugation. The currents can be either the source currents of the electric field $\mathbf{E}_\textrm{inc}$ or, like in our case, the loss currents induced by the external field. For the latter case of extinguished power, equation~\eqref{eq:scalarprod2} will provide negative values.

To facilitate the calculations, we replace the x-y-plane by the complex plane. For our specific geometry this results in complex quantities in polar form instead of vectors containing trigonometric functions.  Two things have to be considered when doing so. First, we write the field and the currents as purely real valued quantities to avoid confusion due to different meanings of the imaginary part. Hence, the complex conjugation in equation~\eqref{eq:scalarprod2} is obsolete. Still, our observables become complex, now due to our choice of the complex plane. Second, as a consequence the definition of the scalar product in equation \eqref{eq:scalarprod2} has to be changed to:

\begin{equation}
P_{\textrm{ext},n} = \frac{1}{2} \textrm{Re} \int_V \mathbf{j}\left(\mathbf{r}\right) \cdot \mathbf{E}^*_\textrm{inc}\left(\mathbf{r}\right) \, \textrm{d}V\,
\label{eq:scalarprod3} \, ,
\end{equation}

\begin{figure}[htb]
\centering
\includegraphics[width=0.4\textwidth]{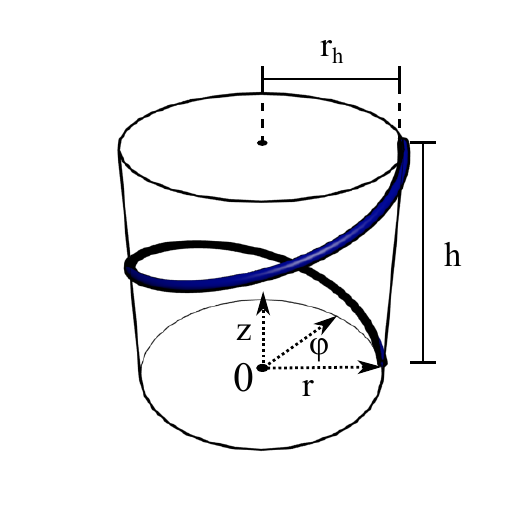}
\caption{Geometry of a single pitch of an 1D helix with radius $r_h$ and pitch height $h$ in a cylindrical coordinate system.}
\label{fig:helixGeom}
\end{figure}

Let us start with the calculation. The transversal excitation field propagates along the z axis and has components pointing in the complex plane only. In our calculation scheme the circularly polarized light $E_x = cos (kz)$ and $E_y = \pm sin(kz)$ read as:
\begin{equation}
E_\varphi(z) = E_0 \cdot e^{\pm ikz}\quad \, .
\label{eq:circFields}
\end{equation}
with $k=2\pi/\lambda$ being the k-vector of the free space excitation light with a wavelength of $\lambda$ and the plus (minus) sign denoting RCP (LCP) light.  

The plasmonic helix stretches along the z axis as depicted in Fig.~\ref{fig:helixGeom}. It consists of a one-dimensional wire of length $L$ coiled up to $m$ turns, with $r_h$ being the helix radius and $h$ being the height of a single pitch. The overall height of the helix is $H=m\cdot h$, with a wire length of $L = m \sqrt{4\pi^2 r_h^2 + h^2}$. The lower end of the wire starts at the point $(x,y,z)=(r_h,0,0)$. Because of the transversality of the incident field, only the in-plane-component of the helix position vector $\vec{h}$ is needed:
\begin{equation}
h_\varphi = r_h e^{\pm ik_h z } \, ,\, z \in [0,H] \,.
\end{equation}
Here, we introduced the helix wave vector $k_h =2\pi/h$ (cf. Fig. 2 c) in the main manuscript). The sign of the exponent again defines the helicity with the plus (minus) sign denoting a right-handed (left-handed) helix.

The direction of the current flow is given by the tangential unit vector at the helix:
\begin{equation}
d_\varphi(z) = \frac{\partial h_\varphi}{\partial z} = \pm i \cdot e^{\pm ik_h z } =  e^{\pm i (k_h z + \pi/2)} \, .
\end{equation}

Next, we need a description of the scalar current density of the wire modes. The standing waves result from the interference of two counter-propagating waves. Instead of mathematically describing the wave functions onto the helical path, we project them onto the z axis. This is justified since the field components are identical in each point of a respective planar wavefront. The helical geometry is accounted for by vectorial overlap of the tangential helix vector with the fields. Please remember that the product of the tangent with the current density will change sign due to the positive and negative values of the relative charge distributions. Furthermore, it is important to note, that the current density $\mathbf{j}\left(\mathbf{r}\right)$ turns into a current $j(z)$ upon 1D projection having the physical unit $[j(z)] = 1 A$.

Standing waves evolve for multiples of $\lambda/2$ onto the wire length. Here, this is true for the fitting of the projection of the effective wavelength onto the height of the helix: $\lambda^{z}_\textrm{eff} = 2 m h/ n$. We therefore define the corresponding wave vector $k_{FP} = n\pi/(mh) = 2\pi/\lambda^{z}_\textrm{eff}$. Finally, we need to ensure the boundary conditions of vanishing currents at the helix ends $ j(H)= j(0) =0$, leading to:
\begin{equation}
j(z) = j(z)_++j(z)_- = \frac{j_0}{2}\cdot\left( e^{ik_{FP}z}-e^{-ik_{FP}z}\right)
\end{equation}

The current is a real-valued sine function as required in the discussion above. W.l.o.g. we now choose the helix to be right-handed. The complete description of the current is then:
\begin{equation}
j_\varphi(z) =  d_\varphi(z)\cdot(j(z)_++j(z)_-) 
= \frac{j_0}{2}\,\cdot \left( e^{i (k_h z + k_{FP}z )}-e^{i (k_h z - k_{FP}z )}\right)e^{i\pi/2}\\
\end{equation}

The extinction power can then be written as:
\begin{align}
P_{\textrm{ext},n} = \frac{1}{2} \textrm{Re} \int_0^{m\, h} j_\varphi(z) \cdot E^*_\varphi(z) e^{i\pi/2}\, \textrm{d}z \, ,
\label{eq:fullCoupling}
\end{align}
where the factor $e^{i\pi/2}$ is the phase shift that has to be introduced due to the wire mode being excited at resonance. Finally, we calculate the extinction power according to:

\begin{align}
P_\textrm{ext} &= \frac{1}{4} \textrm{Re} \int_0^{m\, h} j_0\,\cdot \left( e^{i ( k_h + k_{FP})z}-e^{i ( k_h  - k_{FP})z}\right) \cdot e^{i\pi/2} \cdot E_0 \cdot e^{\mp ikz} \cdot e^{i\pi/2} \, \textrm{d}z \\
&= \frac{1}{4} \int_0^{m\, h} E_0\,j_0\, \textrm{Re} \left[\left( e^{i (\mp k + k_h + k_{FP})z}-e^{i (\mp k + k_h  - k_{FP})z}\right) \cdot (-1) \right] \textrm{d}z \\
&= \pm\frac{1}{4}E_0\,j_0\,  \int_0^{m\, h}  \cos((k \mp k_h  - k_{FP})z) - \cos((k \mp k_h + k_{FP})z) \, \textrm{d}z \\
&= \pm\frac{1}{4}E_0\,j_0\, m h \left[ \textrm{sinc}((k \mp k_h  - k_{FP})mh) - \textrm{sinc}((k \mp k_h + k_{FP})mh)\right]\, ,
\end{align} 

where we used the identity $\cos(-x) = \cos(x)$ and re-sorted the summands. The same result is obtained for real-valued vectorial trigonometric functions when using the sum formulas.

To calculate the numerical values of the overlap integral one needs to know the free space wave length $\lambda$, that excites the FP-mode on the helix with a wavelength of $\lambda_\text{eff}$.  As required by construction the extinguished power $P_\textrm{ext}$ provides negative values for efficient excitation of the induced FP currents. For clarity we plotted the absolute value in Figure 4 of the main manuscript. 

\section{Numerical modeling}

\begin{figure}[htp]
\centering
\includegraphics[width=\textwidth]{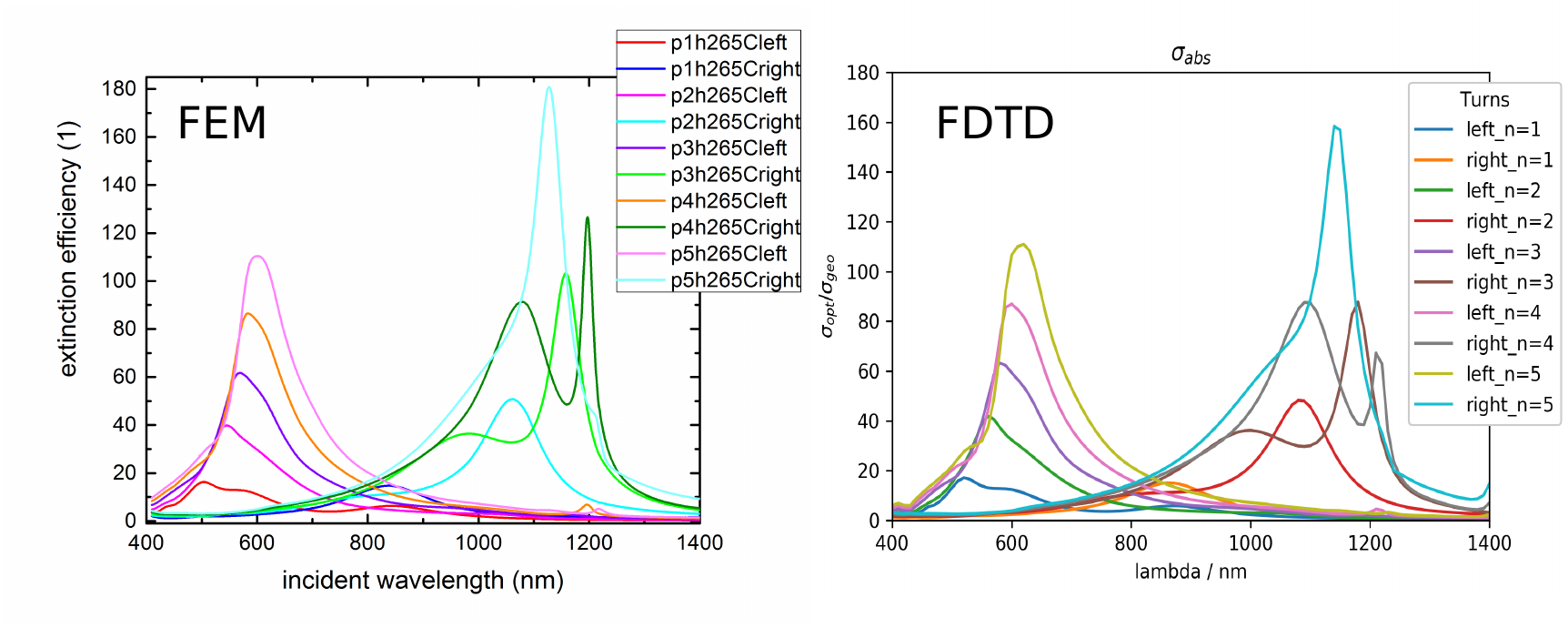}
\caption{\textbf{Comparison between FEM and FDTD numerical results} The display the extinction efficiency of silver helices with 1 -- 5 pitches under left and right circular polarized plane wave illumination.}
\label{fig:FDTDFEM}
\end{figure}

Full field electrodynamic modeling was performed using a finite element solver (Comsol Multiphysics) and a finite difference time domain software (Lumerical Solutions). The modeled geometry in both cases was a single helix surrounded by an air domain and perfectly matched layers (PML) both scaled with the incident wavelength. Both techniques showed good agreement of the calculated spectra, cf. Figure \ref{fig:FDTDFEM}. Here, the modeled helices had a radius of 75\,nm, a pitch height of 265\,nm and a wire diameter of 32\,nm. Slight differences in the spectra can be attributed to discretization issues in FDTD. The cubic mesh elements coarsen the surfaces and thereby slightly increase scattering losses. Furthermore, the staircase error in FDTD causes a minor red shift of resonances. These issues can be addressed by increasing the mesh resolution at the expense of the computational efforts. 
The same discretization issues apply for surface charge distributions which we therefore plotted using the FEM software. In FEM the tetragonal meshing provides the best resolution of curved surfaces. Furthermore, the definition of E and B at the same points in space avoids the artificial hotspots which are typically observed on curved surfaces for FDTD (cf. Figure \ref{fig:dnear} b)). Therefore, all numerical spectra and surface charge plots shown in the main manuscript were calculated using FEM. 

\begin{figure}[h!]
\centering
\includegraphics[width=0.85\textwidth]{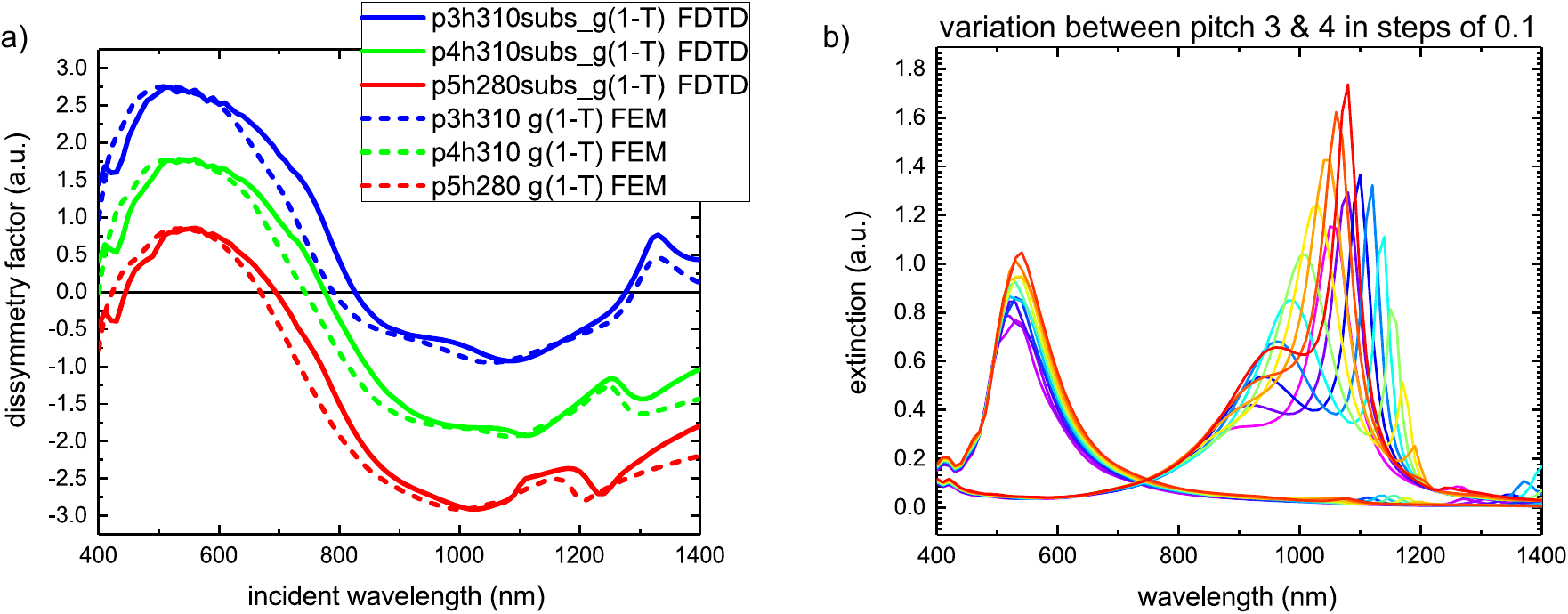}
\caption{\textbf{Further modeling} a) comparison of numerically calculated dissymetry factors with and without substrate, b) variation of pitch number visualizing the envelope provided by the helical design tool.}
\label{fig:var}
\end{figure}

After validating the models, FDTD was employed for a helix onto an ITO-coated glass substrate. Due to the high aspect ratios of the helices, the substrate showed almost no influence on the observed spectral features. The absorption of ITO and glass in the visible range is negligible and, hence, was the difference in dissymmetry (cf Figure \ref{fig:var} a) for the direct comparison of dissymmetry factors). For performance reasons, all further simulations were carried out without substrate.

In the modeling, large parameter sets were tested including radii of helix and wire, height and number of pitches as well as the helix material. These calculations fully support the theoretical considerations. Furthermore, the pitch numbers were continuously increased between their integer values. Figure \ref{fig:var} b) shows the results for varying the pitch number between 3 and 4 in steps of 0.1. Here, the change in the lengths shifts the underlying Fabry-Perot eigenmodes and thereby visualizes the excitation condition derived in the preceding subsection.

\bibliography{silverHelix}